\documentclass[prl,twocolumn]{revtex4}
\usepackage[T1]{fontenc}
\usepackage[latin1]{inputenc}
\usepackage{amsmath}
\usepackage{graphics}
\usepackage{graphicx}
\usepackage{dcolumn}
\usepackage {epsfig}
\makeatother

\providecommand{\LyX}{L\kern-.1667em\lower.25em\hbox{Y}\kern-.125emX\@}

\newcommand{\be}{\begin{equation}}
\newcommand{\ee}{\end{equation}}
\newcommand{\ba}{\begin{eqnarray}}
\newcommand{\ea}{\end{eqnarray}}

\begin{document}

\title[Short Title]{Noncommutative solitons and
instantons\footnote{Talk given at the
{\it XXIV Encontro Nacional de F\'\i sica de Part\'\i culas e
Campos, Caxambu, MG, 30/09-4/10/2003.}}}

\author{Fidel A. Schaposnik}

\email{fidel@athos.fisica.unlp.edu.ar} \affiliation{Departamento
de F\'\i sica, Universidad Nacional de La Plata\\Comisi\'on de
Investigaciones Cient\'\i ficas, Buenos Aires\\ C.C. 67 1900 La
Plata, Argentina}

\begin{abstract}
I review in this talk different approaches to the construction of
vortex and instanton solutions in noncommutative field theories.
\end{abstract}

\maketitle

\section{Introduction}
The development of Noncommutative Quantum Field Theories has a
long story that starts with Heisenberg observation (in a letter he
wrote to Peierls in the late 1930 \cite{P}) on the possibility of
introducing {\it uncertainty relations for coordinates}, as a way
to avoid singularities of the electron self-energy. Peierls
eventually made use of these ideas in work related to the Landau
level problem. Heisenberg also commented on this posibility to
Pauli who, in turn, involved Oppenheimer in the discussion
\cite{O}. It was finally Hartland Snyder, an student of
Openheimer, who published the first paper on {\it Quantized Space
Time} \cite{S}. Almost immediately C.N.~Yang reacted to this paper
publishing a letter to the Editor of the Physical Review \cite{Y}
where he extended Snyder treatment to the case of curved space (in
particular de Sitter space). In 1948 Moyal addressed to the
problem using Wigner phase-space distribution functions and he
introduced what is now known as the Moyal star product, a
noncommutative associative product, in order to discuss the
mathematical structure of quantum mechanics \cite{M}.

The contemporary success of the renormalization program shadowed
these ideas for a while. But mathematicians, Connes and
collaborators in particular, made important advances  in the 1980,
in a field today
known as noncommutative geometry \cite{C}. The physical
applications of these ideas were mainly centered in problems
related to the standard model until Connes, Douglas and
Schwartz observed that noncommutative geometry arises as a
possible scenario for certain low energy limits of string theory
and M-theory \cite{CDS}. Afterwards, Seiberg and Witten \cite{SW}
identified limits in which the entire string dynamics can be
described in terms of noncommutative (Moyal deformed) Yang-Mills
theory. Since then, 1300 papers (not including the present one)
appeared in the
{\bf arXiv} dealing with different applications of noncommutative
theories in physical problems.

Many of these recent developments, including Seiberg-Witten work,
were triggered in part by the construction of noncommutative
instantons \cite{NS} and solitons \cite{GMS}, solutions to the
classical equations of motion or BPS equations of noncommutative
 theories. The
present talk deals, precisely, with the construction of vortex solutions
for the noncommutative version of the Abelian Higgs model and of
instanton solutions for noncommutative Yang-Mills theory. It covers
work done in collaboration with D.H.~Correa, G.S.~Lozano,
E.F.~Moreno and M.J.~Rodr\'\i guez.

~

The plan of the talk is the following. In the next section I
describe the construction of noncommutative field theories using
the Moyal star product and how this can be connected, in the case
of even dimensional spaces, with a Fock space formulation. The
approach will
allow to turn the more involved non-linear equations of motion
(or BPS equations) in noncommutative space into algebraic
equations which are simpler to analyze. The application of
this technique to the construction of vortex solutions in the
noncommutative Abelian Higgs model is presented
 and finally, in the last section,
instanton solutions to the self-dual equation for a $U(2)$
noncommutative gauge theory are discussed.

\section{The connection between Moyal product of fields and
operator product in Fock space}
Let us call $x^\mu$, $\mu =
1,2,...d$ the coordinates of $d$-dimensional
space-time. Given $\phi(x)$ and $\chi(x)$, two ordinary functions
in $R^d$, their Moyal product is defined as \cite{M}
\begin{eqnarray}
\phi(x)*\chi (x) &=& \left.\exp\left(\frac{i}{2} \theta_{\mu
\nu}\partial^\mu_x \partial^\nu_y \right) \phi (x)\chi(y)
\right\vert_{y=x}
\nonumber\\
&=& \phi(x)\chi (x) + \frac{i}{2} \theta_{\mu \nu}\partial^\mu
\phi (x) \partial^\nu\chi(x) \nonumber\\
&&- \frac{1}{8}
\theta_{\mu\alpha}\theta_{\nu
\beta}\partial_\mu\partial_\alpha\phi(x)
\partial_\nu \partial_\beta\chi(x)
 + \ldots \label{moyal}
\end{eqnarray}
with $\theta_{\mu\nu}$ a constant antisymmetric matrix of rank $2r
\leq d$ and dimensions of $(length)^2$.
One can easily see that (\ref{moyal}) defines  a noncommutative
but associative product,
\be \phi(x)*\left(\vphantom{2^3} \chi(x)* \eta(x) \right) =
\left(\vphantom{2^3} \phi(x)* \chi(x) \right)* \eta(x) \ee
Under certain conditions, integration over $R^d$ of Moyal products
has all
the properties of the the trace (Tr) in matrix calculus,

\be
 \int dx \,\phi(x)*\chi (x) = \int dx\, \chi(x) * \phi(x) =
 \int dx\, \phi(x) \chi(x)
 \label{idem}
\ee
Indeed, identity (\ref{idem}) holds when derivatives of fields vanish
sufficiently rapidly
at infinity, since
\begin{eqnarray} \phi(x)*\chi (x) &=& \phi(x)\chi (x) + \frac{i}{2}
\theta_{\mu
\nu}\partial^\mu \phi (x) \partial^\nu\chi(x) - \nonumber\\
&&\frac{1}{8}
\theta_{\mu\alpha}\theta_{\nu \beta}\partial_\mu\partial_\alpha\phi(x)
\partial_\nu \partial_\beta\chi(x) + \ldots \nonumber\\
\nonumber\\
&=& \phi(x)\chi (x) + \partial_\mu\Lambda_\mu \nonumber
\end{eqnarray}
 One has also in this case
cyclic property of the star product,
\be
 \int dx \,\phi(x)*\chi (x)*\psi(x) =
 \int dx\, \psi(x)*\chi(x) * \phi(x)
\ee
Finally, Leibnitz rule holds
\be
\partial_\mu \left(\phi(x)*\chi (x)\right) = \partial_\mu \phi(x)*\chi (x)
  + \phi(x)*\partial_\mu\chi (x)
\ee

The $*$-commutator,
denoted with $[~,~]$,
\be [\phi,\chi] = \phi(x)*\chi(x) - \chi(x)*\phi(x) \ee
is usually called a Moyal bracket.
If one considers the case in which $\phi$ and $\chi$ correspond to
space-time coordinates $x^\mu$ and $x^\nu$, one has, from
eq.(\ref{moyal}),
\be [x^\mu,x^\nu] = i \theta^{\mu\nu} \label{alg} \ee
This justifies the terminology ``noncommutative space-time'' although in
the Moyal product approach to noncommutative field theories one
takes space as the ordinary one  and it is
through the star multiplication of fields that noncommutativity
enters into play.
For example, the action for a massive self-interacting scalar
field takes, in the noncommutative case, the form
\be
S = \int d^4x \left(\partial_\mu \phi  * \partial^\mu \phi -
\frac{m^2}{2} \phi*\phi - \frac{\lambda}{4!} \phi *\phi*\phi*\phi
\right)
\label{action}
\ee
Note that due to eq.(\ref{idem}), the quadratic part of the action
coincides with the ordinary one (and hence Feynman propagators are the
same for commutative and noncommutative theories). It
is through interactions
that differences arise.

~

We are interested in coupling scalars to gauge fields. Given
 a gauge connection $A_\mu$ and a gauge group element $g \in G$,
 the gauge connection should transform,
under a gauge rotation as
\be
A_\mu^g(x) = g(x) * A_\mu(x) * g^{-1}(x) + \frac{i}{e} g^{-1}(x)*
\partial_\mu g(x)
\ee
Note that even in the $U(1)$ case, due to noncommutative multiplication,
the second term in the r.h.s.
has to
be present in order to have a consistent definition of the curvature.
Also, the expression for $g(x)$ as an exponential
should be understood as
\be
g(x) = \exp_*(i\epsilon(x)) \equiv 1 + i\epsilon(x)  - \frac{1}{2}
\epsilon(x) * \epsilon(x) + \ldots
\ee
Accordingly,  even in the $U(1)$  case
the curvature $F_{\mu\nu} $ necessarily implies
a gauge field commutator,
\be
F_{\mu\nu} = \partial_\mu A_\nu - \partial_\nu A_\mu  -ie\left(
A_\mu*A_\nu - A_\nu*A_\mu\right)
\ee
and then, as  it happens for non Abelian
gauge theories in ordinary space, the field strength $F_{\mu\nu}$
  is not gauge invariant but gauge covariant,
\be
F_{\mu \nu} \to g^{-1}* F_{\mu \nu}*g
\ee
However, due to the trace property of the integral,
 the Maxwell action is gauge invariant,
\be
S = \frac{1}{4}\int d^4x F_{\mu \nu} * F^{\mu \nu}
\ee

As for matter fields, one can write
\begin{eqnarray}
D_\mu^{f} \phi = \partial_\mu + i A_\mu * \phi
~ ~ ~  {\rm ``fundamental"}
\nonumber
\end{eqnarray}
 but also
\begin{eqnarray}
& & D_\mu^ {\bar f} \phi = \partial \mu -i  \phi *A_\mu
~ ~ ~  {\rm ``anti-fundamental"} \nonumber\\
~\nonumber
\\
& & {D}_\mu ^{ad}\phi = \partial_\mu \phi  -ie
(A_\mu*\phi - \phi*A_\mu) ~ ~ ~
 {\rm ``adjoint"  }\nonumber
 \\
\end{eqnarray}

Extending non-Abelian gauge theories
 with generators $t^a$ to the noncommutative case
is problematic. Consider for example the case of $G=SU(N)$.
In the commutative case one has
\begin{eqnarray}
[A_\mu,A_\nu] &=& A_\mu^a  A_\nu^b t^a t^b -
A_\nu^b  A_\mu^a  t^b t^a \nonumber\\
~ \nonumber\\
&= &
A_\mu^a  A_\nu^b \underbrace{\left(t^a t^b - t^b t^a\right)}_{
f_{abc} t_c}\label{aco}
\end{eqnarray}
so that the commutator, and {\it a fortiori} the  field strength,
 take values, as the gauge field itself,
in the Lie algebra of the gauge group.
In contrast, in the noncommutative case the presence of the
star product prevents  to arrange
the commutator as above,
\be
[A_\mu,A_\nu] = A_\mu^a * A_\nu^b t^a t^b -
A_\nu^b * A_\mu^a  t^b t^a \nonumber
\ee
Using
\be
t^a t^b = 2i f_{abc} t^c +  \frac{2}{N} \delta_{ab}
I + 2 d_{abc}t^c
\ee
we see that
\be
F_{\mu\nu} = ^{(1)}\!\!\!F_{\mu\nu}^a t^a + ^{(2)}\!\!\!F_{\mu\nu} I
\ee
and hence $F_{\mu\nu}\not \in
{\cal SU(N)}$. One should instead choose $U(N)$ as gauge group since,
in that
case, no problem of this kind arises.

\section{Noncommutative solitons}
In order to understand the difficulties and richness one encounters
when searching for noncommutative solitons, let us disregard the
kinetic energy term in action (\ref{action}) and just consider
 the scalar potential,

\be V[\phi *  \phi] =  \frac{1}{2} m^2 \phi * \phi -
\frac{\lambda}{4} \phi*\phi*\phi*\phi \label{pot}
\ee
The   equation for its  extrema  is
\begin{equation}
m^2 \phi - \lambda \phi*\phi*\phi = 0
\end{equation}
or, with the shift
$
\phi \to ({m}/{\sqrt \lambda}) \phi
$,
\begin{equation}
\phi(x)*\phi(x)*\phi(x) = \phi(x)  \label{chiste}
\end{equation}
To find a solution, consider a function
 $\phi_0(x)$ such that
\begin{equation}
 \phi_0(x) * \phi_0(x) =  \phi_0(x)
 \label{2}
\end{equation}
which evidently satisfies
(\ref{chiste}).
Although simpler than (\ref{chiste}),
 (\ref{2}) implies, through Moyal star products,  derivatives
 of all orders as it was the case for the original equation. Only
some solutions can be found straightforwardly or with some little work.
For example, in $d=2$ dimensions one finds
\begin{eqnarray}
\phi_0 &=& 0 \nonumber\\
\phi_0 &=& 1 \nonumber\\
\phi_0 &=& \frac{2}{\sqrt\theta}
{\exp\left(-\vec x^2/\theta
\right)}_{~ ~ ~ \overrightarrow{\theta \to
0} ~ ~ ~} \delta^{(2)}(x) \label{primera}
\end{eqnarray}
Already a solution like (\ref{primera}) shows that nontrivial regular
solutions, which
were excluded in the commutative space due to Derrick theorem, can
be found in noncommutative space. The the reason for this is clear:
the presence of the  parameter $\theta$ carrying
 dimensions of $length^2$,   prevents the  Derrick scaling
 analysis leading to the
negative results in ordinary space.

Finding more general solutions needs  new angles of attack.
A very fruitful approach was developed in \cite{GMS} by exploiting
an isomorphism between the algebra of functions with the
noncommutative
Moyal product and the algebra of operators on some Hilbert space. We
shall describe this procedure below in a simple two-dimensional example
(but any even dimensional space can be treated identically).

We then start with  two-dimensional space with complex coordinates
\be
z = \frac{1}{\sqrt{2}}(x^1 + i x^2)\, , \;\;\;\;\;\; \bar z=
\frac{1}{\sqrt{2}}(x^1 - i x^2)
\ee
Changing the coordinate normalization,
\be
\hat a = \frac{1}{\sqrt{2\theta}}(x^1 + i x^2)\, , \;\;\;\;\;\;
\hat  a^\dagger = \frac{1}{\sqrt{2\theta}}(x^1 - i x^2)
\ee
one ends with noncommutative coordinates satisfying
\be
[x^1,x^2] = i \theta \longrightarrow [ \hat a, \hat a^\dagger] = 1
\ee
Then $\hat a$ and $\hat
a^\dagger$ satisfy the algebra of  annihilation and creation
operators.  One then considers a Fock space with a basis $|n\rangle$
provided by the eigenfunctions of the number operator $N$,
\be
 \hat N = \hat{a}^\dagger \hat  a  ~ ~ ~ ~ ~ ~     \hat N |n \rangle = n
|n \rangle
\ee
Note that one can establish a connection between  {$n$}
and the radial variable,
\be
\hat N = {\hat a}^\dagger \hat a \approx \frac{1}{2\theta}( x^2 + y^2) =
\frac{r^2}{2\theta}\, , \;\;\; \theta\ \to 0
\ee
 Configuration
space at infinity then corresponds to $n \to \infty$
in Fock space.
Now, it is very easy to write projectors $P_n$ in Fock space,
\be
P_n = |n\rangle\langle n| \; , \;\;\; P_n^2 = 1
\ee
so that
\be
P_n^3 = P_n
\ee
 which is nothing but the configuration space equation (\ref{chiste})
for the minimum of the potential, but
written in Fock space.
So, we can say that we know a
solution to (\ref{chiste}) in operator form,
\be
O_\phi = |n\rangle \langle n|
\ee
 Now, how does one pass from this
solution in Fock space to the corresponding solution in
configuration space?
The answer is to use the
Weyl connection
which can be summarized as follows:
given a field $\phi(z,\bar z)$ in configuration space, take its Fourier
transform
\be
\tilde \phi(k,\bar k) = \int d^2z \phi(z,\bar z) \exp
\left(\frac{i}{\theta}(\bar k z + k \bar z)
 \right)
 \ee
with variables defined as before,
\be
 z = \frac{1}{\sqrt 2} \left(x^1 + i x^2\right)  \;\;\;\; k =
  \frac{1}{\sqrt 2} \left(k^1 + i k^2\right)
\ee
From it,  define the associated operator
\be
O_\phi(a,a^\dagger)  =
\frac{1}{4\pi^2 \theta}
\int d^2k
\tilde\phi(k,\bar k)
\exp\left(-i\bar k a + k a^\dagger\right)
\ee
Then, one can prove that
\be
\underbrace{O_\phi O_\chi}_{operator~product} =
O_{\!\!{\underbrace{\phi*\chi}_{*\,product}}}
\ee
Hence, the complicated star product of fields in configuration space
becomes just a simple operator product in Fock space.
 As an example of how this connection can be used, consider the
expression for $P_n$ (that
can be found in any textbook on second quantization)
\be
|n \rangle \langle n| = : \frac{1}{m!}a^{\dagger n}
\exp(-a^\dagger a) a^{n} :
\ee
(Here ``$:\,:$'' means  normal ordering)
Then, Weyl connection implies
\begin{eqnarray}
 : \frac{1}{m!}a^{\dagger n}
&&\!\!\!\!\!\!
\!\!\!\exp(-a^\dagger a) a^{n} :  \nonumber\\
&= &\int \frac{d^2k}{4\pi^2} \tilde\phi_0^n(\bar k, k):e^{-i(k_{\bar z} a + k_z
a^\dagger)}: \nonumber
\end{eqnarray}
or, anti-transforming (and using Rodrigues formula)
\be
 \tilde\phi_0^n(\bar k,k) = 2\pi \exp(-k^2/2)
 L_n(k^2/2)
 \ee
where $L_n$ is the Laguerre polynomial of order $n$.
Finally, Fourier transforming this expression, one can write
in configuration space
\be
\phi_0^n(\bar z,z) =
2 (-1)^n \exp\left( \frac{\bar z z}{\theta}\right)
L_n\left(\frac{2\bar z z}{\theta}\right)
\label{ficero}
\ee
In this way, any operator solution in Fock space can be connected with
the corresponding solution in configuration space where
fields are multiplied using the star product. In particular,
a general solution for
the minimum of the potential equation
\be
\phi(\vec x)  = \phi(\vec x) * \phi(\vec x) * \phi(\vec x)
\ee
in $d=2$ space  is then,
\begin{eqnarray}
{\rm in~Fock~space:\hspace{1cm}}
P_\phi &=& \sum \lambda_n |n\rangle \langle n|
 \nonumber\\
{\rm in~configuration~space:\hspace{1.2cm}}\phi & =
&\sum \lambda_n \phi_0^n(\vec x) \nonumber\\
\label{sumas}
\end{eqnarray}
with $ \lambda_n = 0, \pm 1 $ and $\phi_0^n$ given by (\ref{ficero}).

Now,  we want more than solving equations for the extrema of potentials.
We then have to be able   to write kinetic energy terms
in Fock space. To this end, observe that
\be
[a^\dagger, a^n] = -n a^{n-1}
\label{unosd}
\ee
We then see that we can identify
\be
\frac{\partial f}{\partial a} =  - [a^\dagger,f(a)]
\ee
so that derivatives of fields $\phi$  become, in operator language,
\be
\partial_z \phi  \to  -\frac{1}{\sqrt \theta}
[ a^\dagger,O_\phi]\, , \;\;\;\;\;\; \partial_{\bar z}\phi \to
 {\frac{1}{\sqrt \theta} [ a,O_\phi]}
 \ee
and the Lagrangian associated to action (\ref{action}) can be written
 in the form
\be
L = \frac{1}{2} \left([a,O_\phi]^2 + [a^\dagger,O_\phi]^2
\right) - \frac{m^2}{2} O_\phi^2 + \frac{\lambda}{4}O_\phi^4
\ee
A last useful formula for the connection relates integration
in configuration space with trace of operators in Fock space:
\be
{\noindent \int dx dy \phi(x,y) \;\;\;\Rightarrow} \;\;\;
2\pi
\theta \,{\rm Tr}\, O_\phi
\ee
From here on we shall abandon the notation $O_\phi$ for operators
and just write $\phi$ both in configuration and Fock space.

\section{Noncommutative vortices}
The noncommutative version of the  Abelian Higgs Lagrangian
(in the fundamental representation) reads
\be
L = -\frac{1}{4} F_{\mu\nu}* F^{\mu \nu}+ \overline{D_\mu \phi}*
D^\mu\phi - \frac{\lambda}{4}( \phi* \bar \phi - \eta^2)^2
\label{lag}
\ee
Let us briefly review how   vortex solutions
were found in such a model in ordinary space
\cite{NO}-\cite{Bogo}. The energy
for
static, z-independent configurations is,
for the commutative version of the theory,
\be
E = \frac{1}{2} F_{ij}^2 + \overline{D_i\phi}{D_i\phi} +
\frac{\lambda}{4} (|\phi|^2 - \eta^2)^2
\ee
Here $i=1,2$ so one can consider the model in  two
dimensional Euclidean
space with
\be
D_i \phi = \partial_i - i A_i \phi \; , \;\;\; \phi = \phi^1 + i \phi^2
\ee
The Nielsen-Olesen strategy
to find topologically non trivial regular solutions to the
equations of motion of this model in ordinary space
can be summarized in the following
 steps going from the trivial to the vortex solution:
\begin{enumerate}
\item \underline{Trivial}  solution
\be
|\phi |= {\eta}  \; , \;\;\; A_i = {0}
\ee
\item Topologically \underline{non-trivial} but  \underline{singular}
 solution (fluxon)
\be
\phi =  {\eta} \exp(i n\varphi) \; , \;\;\; A_i =
n \partial_i \varphi
\ee
with
\be
\int d^2x \varepsilon_{ij}F_{ij} = 2\pi n
\ee
 but
\be
\varepsilon_{ij}F_{ij} = 2\pi {n\delta^{(2)}(x)}
\ee
\item \underline{Regular}  Nielsen-Olesen vortex solution
\be
\phi =  { f(r)}\exp(i n\varphi) \; , \;\;\; A_i =
 {a(r)} \partial_i \varphi
\ee
with  boundary conditions:
\be
f(0)=a(0)= 0 \; , \;\;\; f(\infty) = \eta \; , a(\infty) = n
\ee

\item Bogomol'nyi bound
\be {\rm if~}
\lambda = 2 \; , \;\;\; E \geq 2\pi n
\ee
whenever  the following \underline{first order}
``Bogo\-mol'\-nyi''
equations hold
\begin{eqnarray}
F_{z \bar z} = \eta^2 - \bar \phi  \phi &~ ~ ~ &
-F_{z \bar z} = \eta^2 - \bar \phi \phi \nonumber\\
D_{\bar z} \phi = 0 ~ ~ ~ ~
 &~ ~ ~ & ~ ~ ~ ~ D_z \phi = 0 \nonumber\\
{\rm Self dual ~ ~ ~ ~ } &~ ~ ~ &
{\rm ~ ~ Anti selfdual}
\end{eqnarray}
\end{enumerate}
Exact solutions of these equations can be easily constructed. Let us
describe
as an example the
 noncommutative selfdual case. One
just copies  the commutative strategy, starting from the ``trivial''
solution that we found in terms of projectors
\begin{eqnarray}
 \underbrace{|\phi| = {\eta}}_
{trivial} \hspace{0.2cm}  &\Rightarrow&  \hspace{0.2cm} \phi =
\eta \sum
\underbrace{f_m}_{0, \pm 1}|m\rangle \langle m |
\nonumber\\
 \underbrace{|\phi| = \eta \exp(i\varphi)}_
 {singular}= \eta \frac{z}{|z|}
  \hspace{0.2cm} & \Rightarrow &  \hspace{0.2cm}
\phi = \eta \sum
{f_m} |m\rangle \langle m | \hat a
\nonumber
\\
\underbrace{|\phi|\! = \!f(r) \exp(i\varphi)}_
{regular}= \!f\frac{z}{|z|}
 \hspace{0.2cm} &\Rightarrow&  \hspace{0.2cm}
\phi = \sum
{\bar f_m}|m\rangle \langle m | \hat a\nonumber\\
\label{numb}
\end{eqnarray}
Note that in the second and third lines we have used the identification
$z \to (1/\sqrt \theta) \hat a$. The difference between these two
formul\ae is that
in the second the coefficients are $f_m=\pm 1$ while in the third
one the $f_{\bar m}$  should be
adjusted using the eqs. of motion and boundary conditions.

Of course (\ref{numb}) should be
accompanied by
 a consistent   ansatz for the gauge field,
\be
\hat A_z = \sum {\bar{e}_n } {\hat a}^\dagger|m\rangle \langle m|
\ee
Differential equations (eqs. of motion or Bogomol'nyi eqs.) become
algebraic recurrence relations which can be easily solved. For example, in the
selfdual case
one has from Bogomol'nyi equations
\begin{eqnarray}
&&\sqrt{(n+2)}(f_{n+1}-f_{n})-e_n f_{n+1}=0  \nonumber\\
&& 2\sqrt{(n+1)} e_{n-1}-e^2_{n-1}- 2\sqrt{(n+2)} e_{n}-e^2_{n}
\nonumber\\
& & ~ \hspace{5cm}=-
\theta \eta^2 (f_n^2-1) \nonumber
\end{eqnarray}
The appropriate
condition at infinity ($ |z| \to \infty $) was, in
configuration
space
$f(|z|) \to 1 $. It
 translates to
$f_n \to 1 $ for $ n \to \infty$. Then, using  $f_0$ as a shooting parameter, one determines
 $f_1, f_2, \ldots$ and from them one  computes the magnetic
field, the flux,  the energy, from the expressions
\begin{eqnarray}
&&B(r) = 2 \eta^2 \sum_{n=0}^{\infty}\!(-1)^n \left(1 -
f_n^2\right) \exp\left({-\frac{r^2}{\theta}}\right)
 L_n(\frac{2r^2}{\theta})
\nonumber\\
&& \Phi = 2 \pi \theta {\rm Tr} \hat B = 2\pi
\nonumber\\
&&E = 2\pi \eta^2 \end{eqnarray}
For small $\theta$ one re-obtains the Nielsen-Olesen
regular vortex solution.
 Exploring the whole range of $\theta \eta^2$,
the dimensionless parameter
governing noncommutativity, one finds that
the vortex solution with $+1$ units of magnetic flux  exists  in
all  the  $\theta$ range. As an example, we show in Figure 1 the magnetic field of a
self-dual vortex
with $N=1$ units of magnetic flux, for different values of $\theta$.
We see that the solution approaches smoothly the commutative ($\theta = 0$)
limit.

\begin{figure}
\centerline{
\psfig{figure=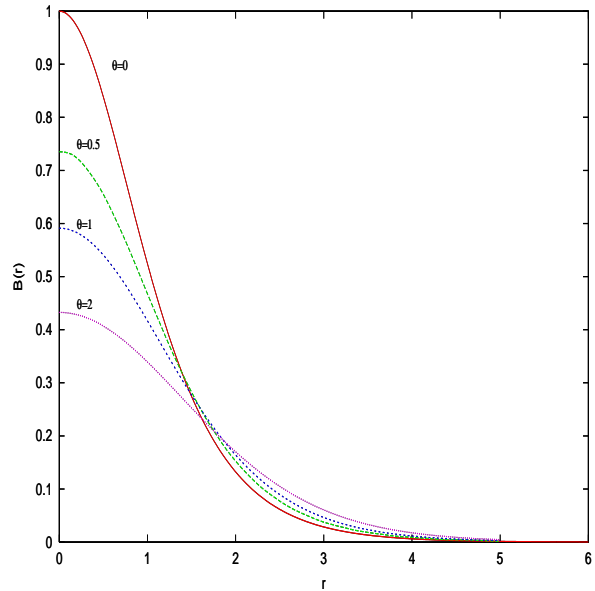,height=8cm,width=8cm,angle=-90}}
\smallskip
\caption{ Magnetic field of a self-dual vortex as a function of the radial
coordinate (in units of $\eta$) for different values of the
anticommuting parameter $\theta$ (in units of $\eta^2$). The curve
for $\theta = 0$ coincides with that of the
Nielsen-Olesen vortex in ordinary space.
\label{fig1} }
\end{figure}

In the commutative case, anti-selfdual solutions can be trivially
obtained from selfdual ones by making $B \to -B$, $\phi \to \bar\phi$.
Now, the presence of  the noncommutative parameter $\theta$,
breaks parity and the moduli space for positive
and negative magnetic flux vortices differs drastically.
One has then to carefully study this issue in all regimes, not only
for $\lambda = \lambda_{BPS}$ but also for
$\lambda
\ne \lambda_{BPS}$, when Bogomol'nyi equations do not hold
and  the second order equations
of motion should be analyzed. A summary of results which are obtained
 is (see details in (\cite{LMS1},\cite{LMS2},\cite{LMRS}):

~

\begin{itemize}
\item {\bf Positive flux}
\begin{enumerate}
\item  {There are BPS and non-BPS solutions in
the whole range of
$\eta^2\theta$. Their energy and magnetic flux are:

\noindent For BPS solutions
\begin{eqnarray}
E_{BPS} &=& 2\pi \eta^2 N \; , \;\;\;    \Phi = 2 \pi N
\nonumber\\
N &=& 1,2,\ldots
\end{eqnarray}

 For non-BPS solutions,
\begin{eqnarray}
E_{non-BPS} &>& 2\pi \eta^2 N \; , \;\;\;    \Phi = 2 \pi N
\nonumber\\
N &=& 1,2,\ldots
\end{eqnarray}
\item For $\eta^2\theta \to 0 $   solutions
 become,
smoothly, the known regular solutions of the commutative case.
\item
In the non-BPS case, the energy of an $N=2$
vortex compared to that of two $N=1$ vortices
is a function of $\theta$.}
\end{enumerate}
As in the commutative case, if one compares the energy of an $N=2$
vortex  to that of two $N=1$ vortices  as a function of
$\lambda$ one finds that for $\lambda> \lambda_{BPS}$ $N>1$ vortices are
unstable (vortices repel) while for  $\lambda < \lambda_{BPS}$ they
attract.

\item {\bf Negative flux}
\begin{enumerate}
\item
 BPS  solutions only exist
in a finite range:
\[0 \leq \eta^2\theta \leq 1\]
Their energy and magnetic flux are:
\begin{eqnarray}
E_{BPS} &=& 2\pi \eta^2 N \; , \;\;\;    \Phi = 2 \pi N  \nonumber\\
N &=& 1,2,\ldots
\end{eqnarray}
\item
When $\eta^2\theta = 1$ the BPS solution becomes a {fluxon},
a configuration which is regular only in the noncommutative
case. The magnetic field of a typical fluxon solution takes the form
\be
B
\sim \frac{1}{\sqrt \theta}{\exp (r^2/\theta)}_
{~ ~ ~  \overrightarrow{{\theta} \to 0}} ~ ~
\delta^{(2)}(x) ~
\ee

\item
There exist non-BPS solutions in the
whole range of $\theta$ but
\begin{enumerate}
\item Only for $\theta <1$
they are smooth deformations of the
commutative ones.
\item For   $\theta \to 1$ they
tend to the fluxon BPS solution.
\item For  $\theta >1$ they coincide
with the non-BPS fluxon solution.
\end{enumerate}
\end{enumerate}
\end{itemize}

\section{Noncommutative instantons}
The well-honored  instanton equation
\be
F_{\mu \nu} = \pm \tilde F_{\mu\nu}
\label{instanton}
\ee
was studied in the noncommutative case by
Nekrasov and Schwarz \cite{NS} who showed that
even in the $U(1)$ case one can find nontrivial instantons. The
approach followed in that work was the extension of the
ADHM construction, successfully applied to the systematic construction
of instantons in ordinary space, to the noncommutative case. This
and other
approaches were discussed
in \cite{i1}-\cite{in}. Here we shall describe the methods developed in
\cite{Cor1},\cite{Cor2}.

We work in  four dimensional space where one can always choose
\begin{eqnarray}
\theta_{12} = - \theta_{21} = \theta_1 \nonumber\\
\theta_{34} = - \theta_{43} = \theta_2 \nonumber\\
{\rm all ~other~} \theta's = 0 \nonumber
\end{eqnarray}
We define dual tensors as
\be
\tilde{F}_{\mu\nu} = \frac{1}{2} \sqrt g \, \varepsilon_{\mu\nu\alpha\beta}
F^{\alpha\beta}
\ee
with $g$ the determinant of the metric.

In order to work in Fock space as we did in the case of
noncommutative vortices, we now need
two pairs of creation
annihilation operators,
\begin{eqnarray}
x^1 \pm i x^2 ~ ~ ~ &{\Rightarrow}& ~ ~ ~ {\hat a}_1, \;\;
{\hat a}^{\;\dagger}_1 \nonumber\\
x^3 \pm i x^4 ~ ~ ~ &{\Rightarrow}& ~ ~ ~ {\hat a}_2, \;\;
{\hat a}^{\;\dagger}_2 \nonumber
\end{eqnarray}
and  the
Fock vacuum will be denoted as $| 0 0 \rangle$. Concerning
projectors  the connection with configuration space takes the form
\begin{eqnarray}
|n_1 n_2\rangle \langle n_1 n_2| &{\Rightarrow}&
\exp\left( -r_1^2/\theta_1 -  r_2^2/\theta_2\right)
L_{n_1}\left(2r_1^2/\theta_1\right)\times \nonumber\\
&& L_{n_2}\left(2r_2^2/\theta_2\right)
\label{suis}
\end{eqnarray}
Finally, note that  the gauge group $SU(2)$ (for which ordinary
instantons were originally constructed) should be replaced
 by $U(2)$ so that
\be
A_\mu = A_\mu^a \frac{\sigma^a}{2}
 + A_\mu^4 \frac{I}{2}
\ee

Let us now analyze how the different ansatz leading
to ordinary instantons can be adapted to the noncommutative case.

~

\noindent\underline {1- (Commutative)'t Hooft multi-instanton ansatz}
(1976)
\begin{eqnarray}
&& A_\mu(x) = \tilde{\Sigma}_{\mu\nu} j_\nu
\nonumber\\
&&
\tilde{\Sigma}_{\mu\nu} = \frac{1}{2}\bar\eta_{a\mu\nu}
 \sigma^a \; , \;\;\;
\bar \eta_{a\mu\nu} =
\left\{
\begin{array}{l}
 ~\varepsilon_{a\mu\nu} \; , \;\;\; \mu,\nu \ne 4\\
  ~~\delta_{a\mu} \; , \;\;\; \nu=4\\
 -\delta_{a\nu}\; , \;\;\;\mu=4
\end{array}
\right.
\nonumber
\\
&&
j_\nu = \phi^{-1} \partial_\nu \phi \nonumber
\end{eqnarray}
Here $\sigma^a$ are the Pauli matrices ('t Hooft ansatz
corresponds to an $SU(2)$ gauge theory). With this ansatz,
the instanton selfdual equation becomes
\be
F_{\mu \nu} = \tilde F_{\mu\nu}
~ ~ ~  \Rightarrow  ~ ~ ~
\frac{1}{\phi} \nabla \phi = 0
\label{lapl}
\ee
with
\begin{eqnarray}
\phi = 1 + \sum_{i=1}^N\frac{\lambda_i^2}{(x - c_i)^2} \; , \;\;\;
\nabla \phi = \sum_{i=1}^N\delta^{(4)}(x - c_i)
\label{sesenta}
\end{eqnarray}
The solution corresponds to a regular
 instanton of topological charge $Q=N$.

 ~

\noindent\underline{2- Noncommutative version of 't Hooft  ansatz}

~

The natural way of extending 't Hooft ansatz is to proceed with the
changes
\begin{eqnarray}
&& j_\nu = \phi^{-1} \partial_\nu \phi  ~ ~ ~  \Rightarrow ~ ~ ~
j_\nu = \phi^{-1} * \partial_\nu \phi +
\partial_\nu \phi * \phi^{-1}
\nonumber\\
&& {A_\mu^a \frac{\sigma^a}{2} =
 \bar \Sigma_{\mu\nu} j_\nu}  ~ ~ ~ \Rightarrow  ~ ~ ~
 A_\mu^a \frac{\sigma^a}{2} =
 \bar \Sigma_{\mu\nu} j_\nu
\label{juta}\\
&& A_\mu^4 =
-i \left(\phi^{-1} * \partial_\nu \phi -
\partial_\nu \phi * \phi^{-1}\right)
\label{jita}
\end{eqnarray}
With this we see that the Poisson equation (\ref{sesenta}) for ordinary
instantons changes according to
\begin{eqnarray}
&& \frac{1}{\phi} \nabla \phi = 0 ~ ~ ~  \Rightarrow  ~ ~ ~
\phi^{-1} *\nabla \phi *\phi^{-1} = 0 \nonumber\\
&&  \nabla \phi = \delta^{(4)}(x)
 \Rightarrow  ~ ~ ~
 \nabla \phi =  \frac{\lambda^2}{\theta_1\theta_2}|00\rangle\langle00|
\end{eqnarray}
One then  gets, for the field strengths,
\begin{eqnarray}
 && F = \tilde F + |00\rangle\langle00|
\label{last}
\end{eqnarray}
We see   that the
self-dual equation is not exactly satisfied: the $|00\rangle\langle00|$
term, the analogous to the delta function in the
ordinary case, is not cancelled as it happened
 with the delta function
source for the Poisson equation (\ref{lapl}) in the commutative case.

~

\noindent\underline{3- Noncommutative BPST ($Q=1$) ansatz} (1975)

~

The pioneering Belavin, Polyakov, Schwarz and Tyupkin ansatz
\cite{BPST} leading
to the first $Q=1$ instanton solution was similar to the
't Hooft ansatz except that $ \Sigma_{\mu\nu}$ was used instead
of its dual $\tilde{ \Sigma}_{\mu\nu}$. Its noncommutative extension
can be envisaged as
\begin{eqnarray}
& & {A_\mu^a \frac{\sigma^a}{2} =
 \Sigma_{\mu\nu} j_\nu} ~ ~ ~   \Rightarrow  ~ ~ ~
A_\mu^a \frac{\sigma^a}{2} =
 \Sigma_{\mu\nu} j_\nu
\label{aotro}
\end{eqnarray}
where $j_\nu$ is defined as in the previous ansatz. Concerning $A_\mu^4$, the consistent
ansatz changes due to the use of $\Sigma_{\mu\nu}$ instead of its dual
as in the 't Hooft ansatz.
One needs now,
instead of (\ref{jita}),
\begin{eqnarray}
& & A_\mu^4 =
i \left(\phi^{-1} * \partial_\nu \phi + 3
\partial_\nu \phi * \phi^{-1}\right)
\label{a4}
\end{eqnarray}
With this, one finally has
\begin{eqnarray}
 & & F_{\mu\nu} = \tilde F_{\mu \nu}\; , \;\;\;    Q = S = 1
\end{eqnarray}
but, due to the necessity of the consistent ansatz for the $A_\mu^4$
component, one can see that
\be
F_{\mu\nu} \ne  F_{\mu \nu}^\dagger
\ee
and hence the price one is  paying in order to have a selfdual field
strength is its  non-hermiticity.
Note however that the action and the topological charge are real.

~

\noindent\underline{4- (Commutative) Witten ansatz} (1977)

~

The clue in this ansatz \cite{Witi}
is to reduce the four dimensional problem to
a two dimensional one through an
 axially symmetric N-instanton ansatz. That is, one passes from
 4 dimensional Euclidan space to 2 dimensional space,
($x^1,x^2,x^3,x^4 \to r,t $) but this last  with
a nontrivial
metric  $g^{ij} = r^2 \delta^{ij}, ~ i,j=1,2$.

The axially symmetric ansatz for the gauge field components
is
\begin{eqnarray}
\vec{A}_r  & = &  A_r(r,t) \vec \Omega(\vartheta,\varphi)  \nonumber\\
\vec{A}_t  & = &  A_t(r,t) \vec \Omega(\vartheta,\varphi) \nonumber\\
\vec A_\vartheta &=& \phi_1(r,t) \partial_\vartheta \vec
\Omega(\vartheta,\varphi)  \nonumber \\
&+ &\left(1 + \phi_2(r,t)\right) \vec \Omega (\vartheta,\varphi)
 \wedge \partial_\vartheta \vec \Omega (\vartheta,\varphi)
\nonumber\\
\vec A_\varphi &=& \phi_1(r,t) \partial_\varphi
 \vec \Omega(\vartheta,\varphi)
\nonumber\\
& +&
\left(1 + \phi_2(r,t)\right) \vec
\Omega (\vartheta,\varphi) \wedge \partial_\varphi \vec \Omega
(\vartheta,\varphi)
\label{ansatz}
\end{eqnarray}
with
\begin{equation}
\vec \Omega(\vartheta, \varphi)  = \left(
\begin{array}{c}
\sin \vartheta \cos \varphi \\
\sin \vartheta \sin \varphi \\ \cos \vartheta
\end{array}
\right)
\ee

With this ansatz, the selfduality instanton equation (\ref{instanton})
becomes a pair of BPS equations for vortices in curved space
\begin{eqnarray}
F_{\mu\nu} = \tilde F_{\mu\nu} &\to &
\left\{
\begin{array}{l}
\frac{1}{\sqrt g}F_{z \bar z} = |\phi|^2 - 1\\
D_z \phi = 0
\end{array}
\right.
\end{eqnarray}
where $\phi = \phi_1 + i \phi_2$ and $z = t + i r$.
Solving these
  BPS vortex equations then reduces to finding the
  solution of a Liouville equation. In this way an
exact axially symmetry N-instanton solution was constructed in
\cite{Witi} for the (commutative)$SU(2)$ theory.

~

\noindent 4-\underline{Noncommutative version of Witten ansatz}

~

To proceed, one needs  a noncommutative setting for curved
2-dimensional space, where $\theta$ can in principle depend on $x$,
\be
[x^i,x^j] =  \theta^{ij}(x)
\ee
Now, handling such a commutator is not trivial since
not all functions $\theta_{ij}(x)$ will guarantee a
noncommutative but associative product.

One can see, however, that
associativity can be achieved whenever
\be
\nabla_i \theta^{ij} = 0
\ee

In the present 2 dimensional case, these equations have as solution
\be
\theta_{ij} = \theta_0 \frac{\varepsilon_{ij}}{\sqrt g}
\ee
with $\theta_0$ a constant. Then, given the metric in
which the instanton problem with axial symmetry reduces to
a vortex problem we see that an associative  noncommutative
product should take the form
\be
[r,t] = r^2 \theta_0 \; ; \;\;\; all~other ~ [.,.]=0
\label{rara}
\ee
with now $r$ and $t$ defining the two dimensional variables
in curved space.
A further simplification occurs after the observation that
\be
 r*t - t* r = r^2 \theta_0  ~ ~ ~
\Rightarrow ~ ~ ~ t   * \frac{1}{r} - \frac{1}{r} * t = \theta_0
\ee
Then, calling $y^1 = t$ and $y^2 = 1/r$ we have
instead of (\ref{rara})
the usual flat space Moyal
product and the  Bogomol'nyi equations take the form
\begin{eqnarray}
 \left( 1 - \frac{1}{2} (z + \bar z )^2\right)D_z \phi &=&
 \left( 1 + \frac{1}{2} (z + \bar z )^2\right)D_{\bar z} \phi
\label{cuss} \\
iF_{z \bar z} & = & 1 - \frac{1}{2}[\phi,\bar \phi]_+
\label{cus}\\
iF_{  z \bar z} & = & - \frac{1}{2}[\phi,\bar \phi] \label{cu}
\end{eqnarray}
with $z = y^1 + i y^2$. We can at this point apply the
Fock space method detailed above for constructing vortex solutions.
In the present case, consistency of eqs.(\ref{cuss})-(\ref{cu}) imply
\be
\bar \phi \phi =1
\ee
and hence the only kind of nontrivial ansatz
should lead, in Fock space, to a scalar field of the form
\be
\phi = \sum_{n=0} |n + q\rangle \langle   n|
\ee
where $q$ is some fixed positive integer.
With this, it is easy now to construct a class of solutions analogous
to those previously found for vortices
  in
flat space. It takes the form
\begin{eqnarray}
\phi &=& \sum_{n=0}|n+q\rangle\langle n|  \nonumber\\
A_z &=& -\frac{i}{\sqrt{\theta_0}} \sum_{n=0}^{q-1} \left( \sqrt{n+1}
\right) |n+1\rangle\langle n|\label{solucion2} +\nonumber\\
&+&\frac{i}{\sqrt{\theta_0}} \sum_{n=q} \left( \sqrt{n+1-q} - \sqrt{n+1}
\right) |n+1\rangle\langle n|\nonumber\\
\label{solucion}
\end{eqnarray}
One can trivially verify
 that configurations (\ref{solucion}) satisfy eqs.(\ref{cuss})-(\ref{cu})
 provided
 $\theta_0 = 2$. In
 particular, both the l.h.s. and r.h.s of eq.(\ref{cuss})
 vanish separately.
The field strength associated to our solution reads,
 in Fock space,
\be
i F_{z \bar z} = -\frac{1}{2} \left(
|0\rangle\langle 0| +   \ldots +
|q-1\rangle\langle q-1|
\right) \equiv B
\label{defB}
\ee
or, in the original spherical coordinates
\begin{eqnarray}
\vec F_{tu} &=& B(r) \vec \Omega \nonumber\\
\vec F_{\vartheta\varphi}
&=& B(r)   \sin \vartheta \, \vec \Omega \nonumber\\
F_{tu}^4 &=& B(r) \nonumber\\
 F_{\vartheta\varphi}^4 &=& B(r) \sin \vartheta
\label{instan}
\end{eqnarray}
As before, starting from  (\ref{defB}) for $B$ in Fock space, we can
obtain the explicit form of $B(r)$ in configuration space  in terms
of Laguerre polynomials, using eq.(\ref{suis}).
Concerning the topological charge, it
is then given by
\begin{eqnarray}
Q &=& \frac{1}{32 \pi^2} {\rm tr} \int d^4 x {\varepsilon}^{\mu \nu \alpha \beta}
 F_{\mu\nu} F_{\alpha \beta}\nonumber\\
&=& \frac{1}{\pi}
\int_{-\infty}^{0}du \int_{-\infty}^{\infty} dt B^2 = 2 {\rm Tr} B^2
= \frac{q}{2}
\end{eqnarray}
We thus see that $Q$ can be in principle integer or semi-integer,
 and this for an ansatz
which is formally the same as that proposed in \cite{Witi} for ordinary
space and which yielded in that case to an integer. The origin of this
difference between the commutative and the noncommutative
cases can be traced
back to the
fact that in the former case,  boundary conditions were imposed on the
half-plane and
forced the solution to have an associated integer number.
  In fact, if one plots Witten's
 vortex solution in ordinary space in
the whole $(r,t)$ plane, the magnetic flux has
two peaks and the corresponding
vortex number is even. Then, in order to parallel this treatment in the
noncommutative case one should impose the condition $q = 2 N$.

\vspace{0.3 cm}

\noindent\underline{Acknowledgements}: I would like to thank the
organizers of the XXIV Encontro Nacional de F\'\i sica de Part\'\i culas
 e
Campos at Caxambu for inviting me to deliver a talk at such an
impressive meeting  and for
their warm hospitality.
This work  is partially
supported
by UNLP, CICBA,  ANPCYT,  Argentina.

\end{document}